\documentclass{PoS}

\title{Inner radio structure of compact BAL quasar 1045+352}

\ShortTitle{Radio structure of BAL quasar 1045+352}

\author{\speaker{Magdalena Kunert-Bajraszewska}\\
Toru\'n Centre for Astronomy, N. Copernicus University, 87-100
Toru\'n, Poland\\
        E-mail: \email{magda@astro.uni.torun.pl}}

\author{Marcin P. Gawro\'nski\\
        Toru\'n Centre for Astronomy, N. Copernicus University, 87-100
Toru\'n, Poland\\
        E-mail: \email{motylek@astro.uni.torun.pl}}

\abstract{
The first multifrequency radio 
observations of the very compact BAL quasar, 1045+352, were made using MERLIN 
and the VLBA in a snapshot mode. However, its unusual radio structure was
still very difficult to interpret, indicating a scenario of intermittent
activity or a jet precession. Here, we present some of a new full-track radio 
observations of 1045+352 made with the EVN+MERLIN at 5 GHz. 
The new more sensitive high-resolution observations made possible to trace
the connection between the arcsecond structure and the radio core, and
showed the presence of strong interactions between the jet and the medium of
the host galaxy.  

}

\FullConference{The 9th European VLBI Network Symposium on The role of VLBI in the Golden Age for Radio Astronomy and EVN Users Meeting\\
		 September 23-26, 2008\\
		 Bologna, Italy}

\begin{document}

\section{Introduction}

Broad absorption lines (BALs) are seen in 10-30\% (depends on the
selection criteria) of both the
radio-quiet and radio-loud quasar populations. They are probably caused
by the outflow of gas with high velocities and are a part of the accretion
process. There are two scenarios explaining origin and nature of BAL
quasars. According to first, BAL regions exist in both BAL and non-BAL 
quasars, and the BAL quasars are normal quasars seen along a particular 
line of sight \cite{elvis00}. The second says that 
BALs may be
associated with an evolutionary phase with a large BAL wind covering
fraction, rather than orientation \cite{gregg00}.  
The radio morphologies of radio-loud BAL quasars provide important
additional information about their orientation  and the direction of the
outflow, but only a handful of them have been resolved using radio observations. 
Most of those discovered to date are compact at radio frequencies with either a 
flat or steep spectrum, suggesting a wide range of orientation.

It has been suggested \cite{becker00} that, because of their small sizes,
most of the radio-loud BAL quasars belong to the class of compact radio
sources, named compact steep spectrum (CSS) objects and gigahertz peaked
spectrum (GPS) objects. GPS and CSS sources are considered to be young radio
sources with linear sizes less then 20\,kpc, entirely contained within the 
extent of the host galaxy.

There are only a few radio-loud compact
BALQSOs with resolved structures and only at one frequency
\cite{jiang03,liu08}. The first multifrequency radio observations
of very compact BAL quasar 1045+352 were made by our group \cite{kun07}.
Since the radio structure of 1045+352 appeared to be very complex and
difficult to interpret based on the obtained images, we have made new high
resolution and high sensitive EVN+MERLIN 5\,GHz observations of this source.
Here we present some of the results of the new observations, the
details of which will be discussed in separate paper.

\section{Observations and preliminary results}

1045+352 is a quasar with a redshift of $z=1.604$ \cite{willott02} at
RA=$10^{\rm h}48^{\rm m}34^{\rm s}247$, Dec=$+34^{\rm o}
57^{''}24^{'}99$ (coordinates for J2000 extracted from FIRST).
It is a CSS object and its largest linear size is equal to 4.3\,kpc (${\rm
H_0}$=71${\rm\,km\,s^{-1}\,Mpc^{-1}}$,
$\Omega_{M}$=0.27, $\Omega_{\Lambda}$=0.73).
1045+352 has a very reddened spectrum showing a
high-ionisation C\,IV broad absorption system \cite{willott02}, so the
source has been clasified as a high-ionisation broad absorption line (HiBAL)
quasar.

\begin{figure*}[t]
\centering
\includegraphics[width=7.5cm, height=8.5cm]{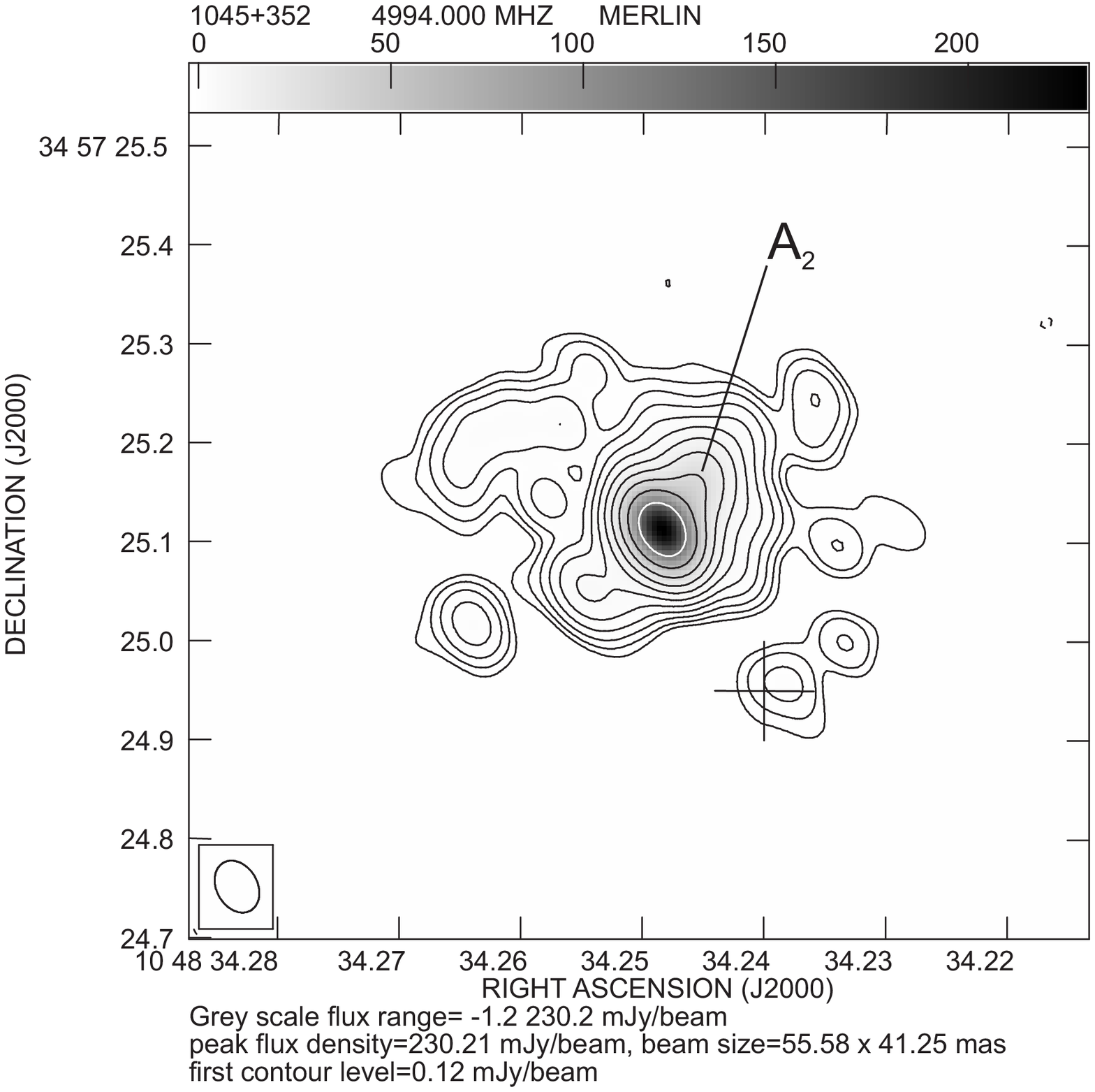}
\includegraphics[width=7.5cm, height=8.5cm]{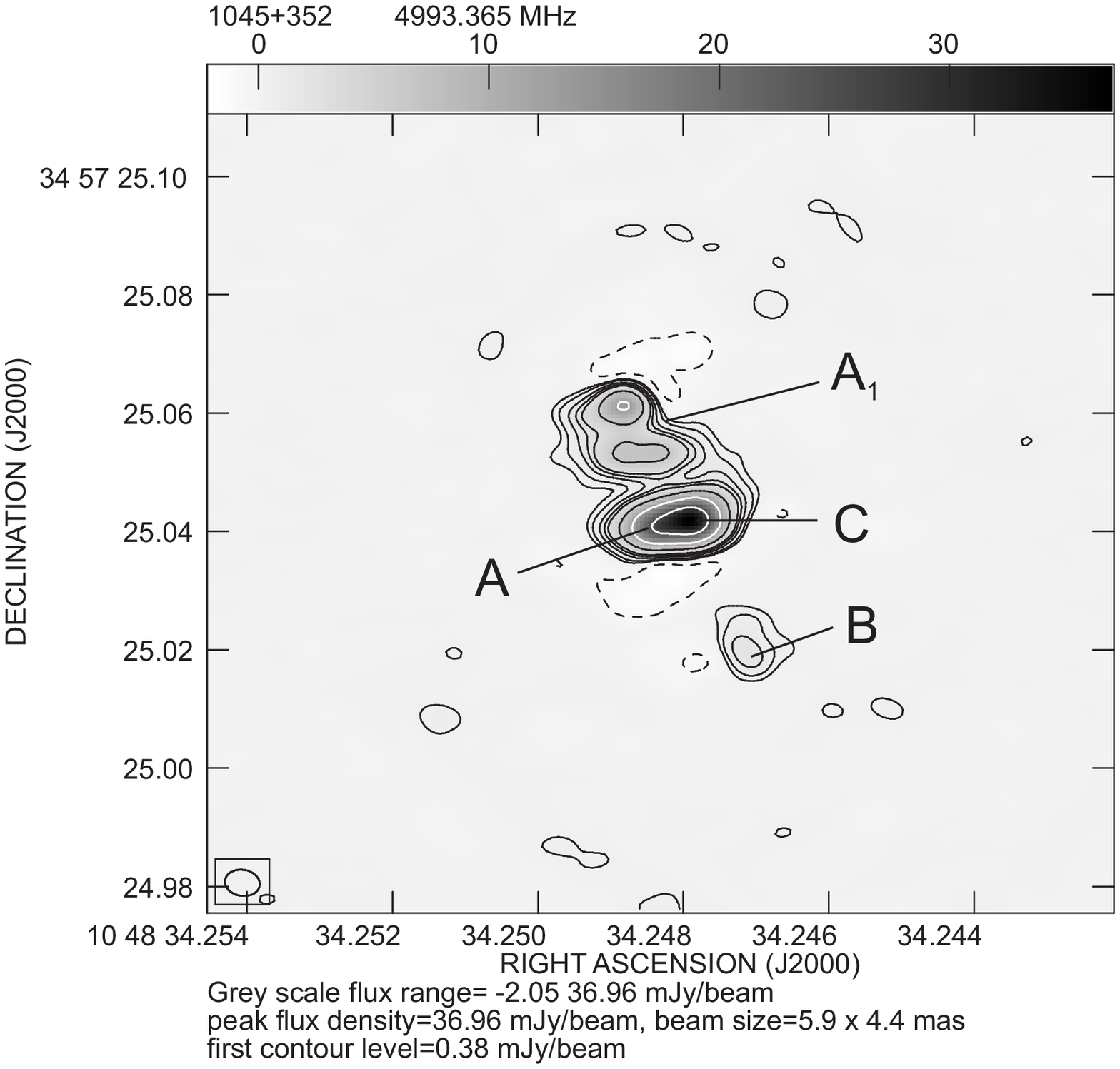}
\caption{The 5\,GHz MERLIN image from
\cite{kun07} and the new 5\,GHz EVN image of 1045+352.
Contours increase
by a factor 2 and the first contour level corresponds to $\approx 3\sigma$.
A cross indicates the position of an optical object found using the SDSS.}
\label{1045+352_maps}
\end{figure*}

1045+352 belongs to the primary sample of 60 candidates for CSS sources
selected from the VLA FIRST catalogue \cite{wbhg97}. Initial observations of
all the candidates were made with MERLIN at 5\,GHz \cite{kun02} and 1.7, 5 and
8.4-GHz VLBA follow-up of 1045+352 was carried out on 13 November 2004 in a 
snapshot mode with phase-referencing \cite{kun07}. 
The new images showed that this source have a complex radio
morphology with a radio jet axis reorientation, that may result from (1) a
merger, (2) a jet precession, or (3) jet-cloud interactions.
However, to indicate one of the above-mentioned scenarios we planned more
sensitive high-resolution EVN+MERLIN 5 GHz observations that was carried out
on 2 June 2007 in a full-track mode with phase-referencing.
The target source scan was interleaved with a scan on a phase reference source
and the total cycle time (target and phase-reference) was
$\sim$7~minutes including telescope drive times, with $\sim$4\,minutes
actually on the target source per cycle.
The whole data reduction process was carried out using standard AIPS
procedures. For the target source, the corresponding
phase-reference source was mapped and the phase errors so determined were
applied to the target source, which were then mapped using a few cycles of
phase self-calibration and imaging. IMAGR was used to produce the final
images separetely for EVN and MERLIN, and finally the
combined EVN+MERLIN image was created. Here, we present preliminary 5\,GHz
EVN image of 1045+352 (Fig.~\ref{1045+352_maps}).

The new 5\,GHz EVN observations showed a jet (indicated as a
A, Fig.~\ref{1045+352_maps}) coming from 
the core (C) in a E/SE direction, and then another jet ($A_{1}$) emerging in a NE
direction. The feature indicated as a B is probably a counter-jet, and a
trace of it was
also visible in a previous lower resolution 1.7-GHz VLBA image \cite{kun07}. 
There is an appreciable ($\sim60^{\rm o}$) misalignment between the axis of the
inner structure (jet A) with respect to the outer structure (jet 
$A_{1}$). 
We suggest the following interpretation of the radio
structure of 1045+352. The radio jet emerging from the core in a E/SE
direction is not able to get through the dense environment and bents
to a NE direction (component $A_{1}$). 
It's continuation ($A_{2}$) is visible on MERLIN 5\,GHz image
(Fig.~\ref{1045+352_maps}).
The weakness of the counter-jet emission is probably caused by
the large beaming. We suspect the angle between the jet axis and the
observer is small, much less than the $30^{\rm o}$ we estimated last time
\cite{kun07}.

Actually there are 17 compact BAL quasars observed with VLBI in a literature
\cite{jiang03,kun07,liu08}[and Montenegro-Montes' talk and proceeding].
About half of them have still unersolved structures even at high
resolution, the other have core-jet structures indicating some orientation
or very complex morphology,
suggesting at least strong interaction with the surrounding medium
\cite{kun07}.

\section{Summary}

1045+352 is a CSS object and a HiBAL quasar with a medium redshift. The new
more sensitive high-resolution EVN 5\,GHz observations revealed much
more radio emission inside the source and confirmed its complicated radio
structure. 
We suggest there are a strong jet-cloud interactions present in that source 
and changing the jet path.
However not all the features visible
on the new radio images can be explained by the interactions with the
surrounding medium. We
suspect there is a jet precession persent in that source, and the detailed
discussion will be presented in a forthcoming paper.

\section*{Acknowledgement}
\noindent
MERLIN is a UK National Facility operated by
the University of Manchester on behalf of STFC.\\
The European VLBI Network is a joint facility of European, Chinese, South
African and other radio astronomy institutes funded by their national
research councils.\\
This work was supported by Polish Ministry of Science and Higher Education
under grant N N203 303635. Partial support for this work was provided by the
UMK under grant no. 410-A.

\end{document}